%% file: main.tex
\documentclass[trackchanges,default,twocolumn]{aastex701}
\usepackage{lipsum}
\usepackage{CJK}
\usepackage{gensymb}

\begin{document}

\title{Measurement of the Half-Light Radius for the Tucana  Dwarf Spheroidal Galaxy}

\begin{CJK*}{UTF8}{bsmi}

\author[orcid=0009-0008-7624-274X,sname='Huang']{Pei-Jun Huang (黃珮君)}
\affiliation{Department of Physics, Brown University}
\email[show]{\url{pei-jun_huang@brown.edu}}

\author[orcid=0000-0003-0751-7312,gname='Ian', sname='Dell\'Antonio']{Ian Dell'Antonio} 
\affiliation{Department of Physics, Brown University}
\email[show]{ian\_dell'antonio@brown.edu}

\author[0009-0005-7997-0652,sname='LaDuca',gname='Philip']{Philip LaDuca}
\affiliation{Department of Physics, Brown University}
\affiliation{Department of Physics, Carnegie Mellon University}
\email{pladuca@andrew.cmu.edu}

\author[0000-0003-0936-7223,sname='Escalante',gname='Zacharias']{Zacharias Escalante}
\affiliation{Department of Physics, Brown University}
\email{zacharias_escalante@brown.edu}

\author[0000-0003-2314-5336,sname='Englert',gname='Anthony']{Anthony Englert}
\affiliation{Department of Physics, Brown University}
\email{anthony_englert@brown.edu}


\begin{abstract} 
    We present a new measurement of the half-light radius of the Tucana Dwarf galaxy, based on the combination of wider, deeper imaging conducted as part of the Local Volume Complete Cluster Survey (LoVoCCS) and HST archival images. We obtain a stellar density profile within the Tucana Dwarf field based on elliptical fitting. After background subtraction, we fit a S\'ersic profile to the density profile to obtain the half-light radius. We measure the half-light radius to be $0.89_{-0.02}^{+0.06}$ arcmin, corresponding to a value of $R_{e} = 225.7_{-5.5}^{+14.4}$ pc (95\% CL). Given the wider-field observations from LoVoCCS used in this analysis, our measurement of the half-light radius represents a $\sim 10 \%$ increased value from the previous results. 
\end{abstract}


\input{sections/intro}
\input{sections/observations}
\input{sections/methods}

\input{sections/results}

\input{sections/discussion}

\begin{acknowledgments}
We acknowledge useful conversations with Alexis Ortega, Kaleb Anderson, and Lawrence Edmond IV.
We would like to especially thank Shenming Fu for his insightful comments and exploratory work.
This work is based on observations made at NSF Cerro Tololo Inter-American Observatory, NSF NOIRLab (NOIRLab Prop. ID 2018A-0308; PI: I. Dell'Antonio), which is managed by the Association of Universities for Research in Astronomy (AURA) under a cooperative agreement with the U.S. National Science Foundation. The Dark Energy Camera data were obtained for the Local Volume Complete Cluster Survey. This work is also based on observations made with the NASA/ESA Hubble Space Telescope, and obtained from the Hubble Legacy Archive, which is a collaboration between the Space Telescope Science Institute (STScI/NASA), the Space Telescope European Coordinating Facility (ST-ECF/ESA) and the Canadian Astronomy Data Centre (CADC/NRC/CSA). The archival Hubble Space Telescope imaging that this work was based on are associated with HST cycle 14 GO-10505 and cycle 18 GO-12273. The computations in this work were run on the Oscar high-performance computing system, supported by the Center for Computation and Visualization at Brown University. 
\end{acknowledgments}

\begin{contribution}

PJH was responsible for data processing, model fitting, and statistical testing, as well as writing and submitting the manuscript as primary author. 
ID came up with the initial research concept, supervised the work, suggested tests to be run, and edited the manuscript. 
PL was responsible for generating astrometry and photometry measurements and contributing to the manuscript.
ZE contributed to the image processing and the manuscript.
AE is the developer of \texttt{lovoccs\_pipe}, which was used for processing DECam observations, and contributed to the manuscript.


\end{contribution}

%
\facilities{HST(WFC/ACS), DECam, CTIO.}

\software{
    Astropy \citep{2013A&A...558A..33A,2018AJ....156..123A,2022ApJ...935..167A},
    Source Extractor \citep{1996A&AS..117..393B}, 
    LSST Science Pipelines \citep{bosch2018overviewlsstimageprocessing, bosch_overview_2019}.}




\bibliography{ref}{}
\bibliographystyle{aasjournalv7}

\end{CJK*}

\end{document}

%% file: sections/intro.tex
\section{Introduction} \label{sec:intro}

Dwarf galaxies are among the most dark matter-dominated galaxies. Studying dwarf galaxies is thus one of the best astrophysical probes of dark matter \citep{battaglia2022stellardynamicsdarkmatter}. The kinematics of visible matter in dwarf galaxies are primarily determined by their dark halos, which can indicate dark matter interactions. Therefore, the measurement of dwarf galaxy mass distributions can provide an important discriminator between dark matter models. Dwarf spheroidal galaxies like Tucana Dwarf are some of the smallest and least luminous galaxies, with mass-to-light ratios of $\frac{M}{L_V} \sim 10^{1-2}$ \citep{Mateo_1998, Walker_2009}. 

Tucana Dwarf is a gas-poor dwarf spheroidal galaxy with no ongoing star formation \citep{Savino_2019}.
First studied and identified as a member of the Local Group by \cite{lavery_1991}, Tucana Dwarf's relatively isolated location is paramount in understanding the evolution of dwarf galaxies away from large spiral galaxy halos. The galaxy experienced a burst of star formation around 13 Gyrs ago, lasting as long as 1 Gyr \citep{Monelli_2010}. A previous study of the stellar metallicities and metallicity gradient in Tucana Dwarf by \cite{fu2024_tucana} found results consistent with predicted gradients from FIRE-2 simulations, featuring bursty star formation that creates stellar population gradients and dark matter cores. 

Previous kinematic studies of the Tucana Dwarf galaxy have measured the systemic velocity and velocity dispersion of the galaxy \citep{saviane_1996,Gregory_2019,Taibi_2020}, yielding differing values for their velocity dispersion and central densities. 
The origin of quiescent, isolated dwarf galaxies, such as Tucana, is not well understood. Previous studies \citep{Sales_2007, Teyssier_2012, Santos_Santos_2023} offer explanations for their presence due to interactions with the Milky Way and/or M31. 

In addition, short-period variables in Tucana Dwarf have been identified and studied. \cite{Bernard_2009} identified 371 variables in Tucana, comprised of a significant fraction ($\sim$17\%) of double-mode RR Lyrae (RRd). By examining RR Lyrae members of Tucana Dwarf, \cite{Bernard_2008} found that the metallicity gradient in Tucana, possessing a higher concentration of metal-rich RR Lyrae variables towards its center, must have appeared early in the galaxy's history.

We can assume that, in a dwarf spheroidal galaxy, the stellar population is in dynamic equilibrium and traces the underlying dark matter-dominated potential \citep{Walker_2009, Chen_2017}. Using the Jeans equations and assuming spherical symmetry, the dark matter mass profile $M(r)$ can be traced from stellar distribution function as
\begin{equation}
    \frac{1}{\nu} \frac{d}{dr} (\nu \bar{v}_r^2) + 2 \frac{\beta \bar{v}_r^2}{r} = -\frac{GM(r)}{r^2}
    \label{eq:jeans}
\end{equation}
where $\nu(r)$ is the stellar number density profile, $\bar{v}_r^2(r)$ is the radial velocity dispersion, and $\beta(r) \equiv 1 - \frac{\bar{v}_\theta^2}{\bar{v}_r^2}$ is the orbital anisotropy \citep{binney_tremaine}. Therefore, obtaining the half-light radius and measuring the stellar number density profile is an important step towards constraining the dark matter mass profile of a dwarf spheroidal galaxy.

The half-light radius of Tucana was most recently measured by \cite{saviane_1996} using imaging from EFOSC2 at the European Southern Observatory's 2.2m telescope. Motivated by the availability of deeper and wider imaging of the Tucana dwarf spheroidal galaxy, conducted as part of the Local Volume Complete Cluster Survey (LoVoCCS), complementary to data from HST, we present an updated measurement of Tucana Dwarf's half-light radius obtained from a S\'ersic-modeled stellar number density profile. 

%% file: sections/observations.tex
\section{Observations}\label{sec:observations}

\subsection{Data acquisition}

\begin{figure}[h!]
    \centering
    \includegraphics[width=0.95\linewidth]{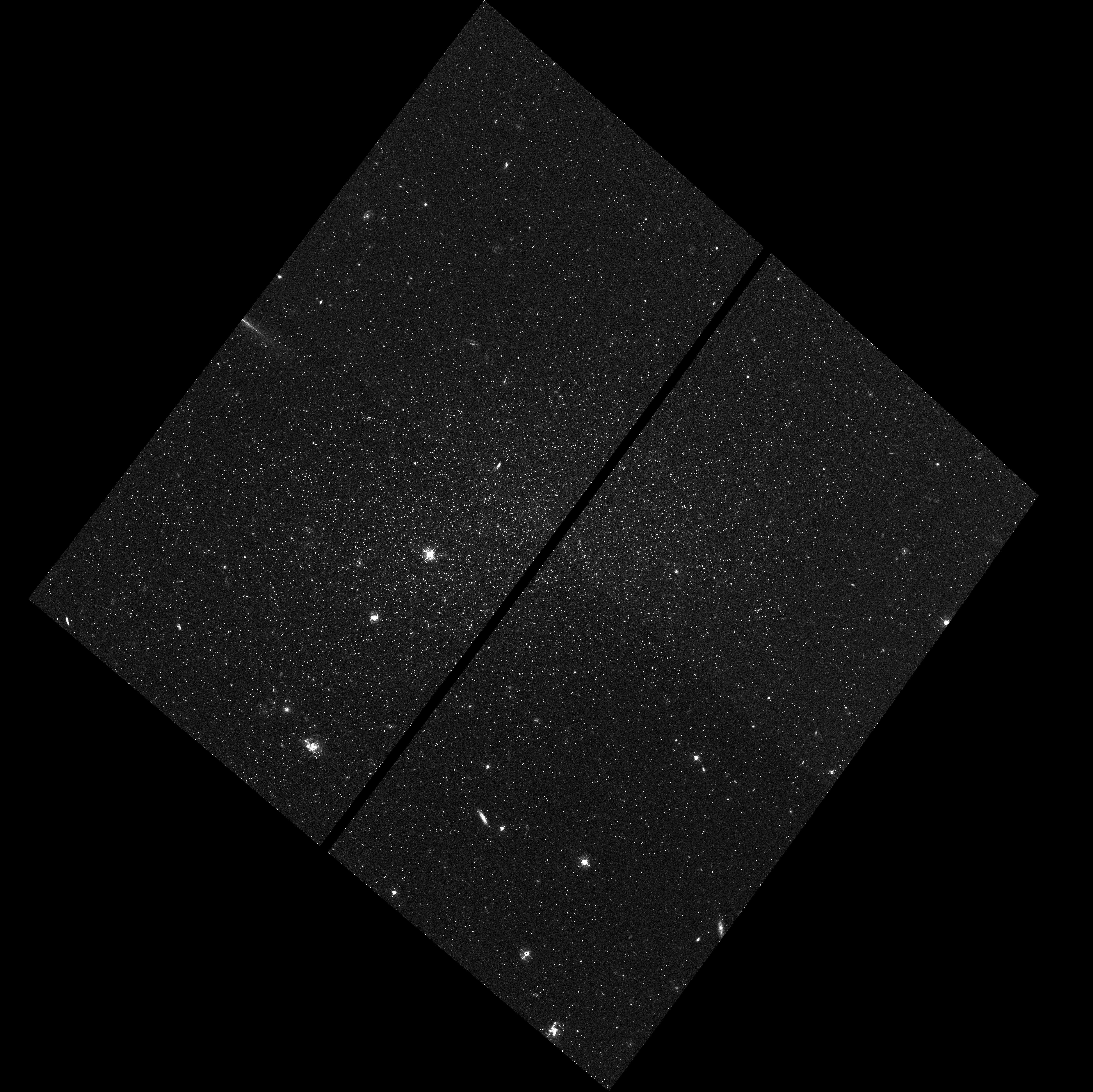}
    \caption{ACS/WFC F475W band image of Tucana Dwarf from HST Proposal GO-10505.}
    \label{fig:hst-img}
\end{figure}

\begin{figure}[h!]
    \centering
    \includegraphics[width=0.95\linewidth]{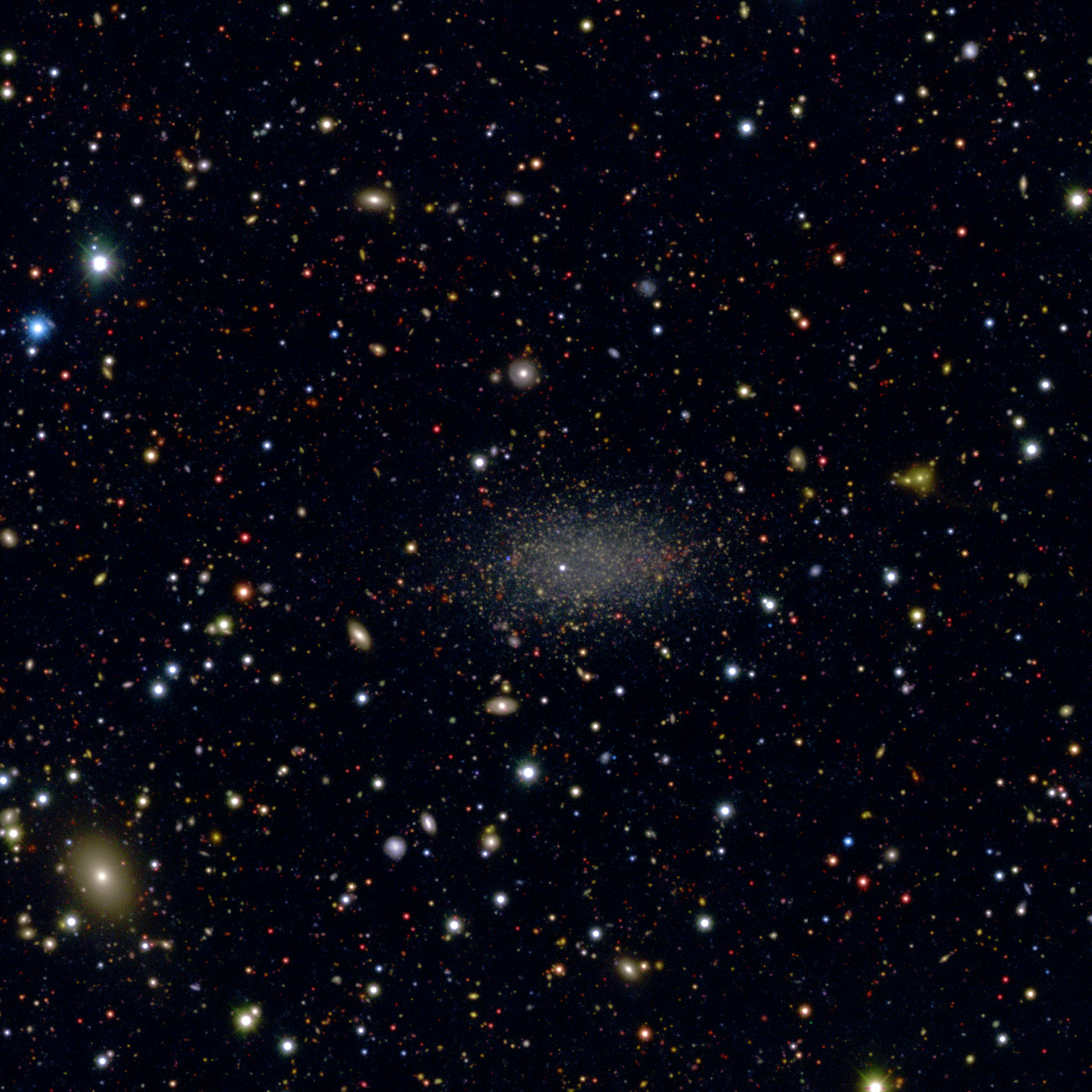}
    \caption{Cropped view of Tucana Dwarf from stacked $i$, $r$, $g$ band images, captured by DECam, processed with the LoVoCCS processing pipeline.}
    \label{fig:lovoccs-img}
\end{figure}

We used HST ACS/WFC F475W and F814W band observation data for the central region of Tucana Dwarf. The F475W band data was obtained from visit 42 of HST Proposal 10505 on April 29, 2006, during HST cycle 14 (GO-10505, PI: Gallart). The F814W band data was obtained from visit 5 of HST Proposal 12273 on May 11, 2011, during HST cycle 18 (GO-12273, PI: van der Marel). The HST images cover a field of view of $\sim 2.5' \times 2.5'$. 
The imaging conducted under Proposal 10505 were associated with the ACS LCID project \citep{Bernard_2008, Bernard_2009, Gallart_2015}, while that of Proposal 12273 are associated with the works of \cite{Sohn_2012, Sohn_2013}. The F475W band image of Tucana Dwarf from Proposal 10505 is shown in Fig. \ref{fig:hst-img}. 

This study utilizes observational data in the $g$, $r$, $i$, $z$ bands taken with the Dark Energy Camera (DECam) from the NSF V\'ictor M. Blanco 4-meter Telescope at Cerro Tololo Inter-American Observatory (CTIO) \citep{flaugher_dark_2015}. The data was acquired as part of the LoVoCCS survey, a study of local ($0.03 < z < 0.12$), X-ray luminous ([$0.1-2.4$ keV]$L_{X500}> 10^{44}$ erg s$^{-1}$) galaxy clusters \citep{fu_lovoccs_2022}. One such cluster, Abell 3921 (A3921), is separated by about 53 arcminutes from the Tucana dwarf, and was thus contained in DECam's 2.2 degree field-of-view when centered on the cluster. A total of 356 raw exposures of the A3921 field were processed using an early version of \texttt{lovoccs\_pipe}\footnote{\url{https://github.com/astroenglert/lovoccs_pipe}} \citep{englert_aas_2025,englert_intracluster_2025}, which reduces DECam observations using \texttt{v26\_0\_0} of the LSST Science Pipelines \citep{LSST_science_pipelines}. Once processed, a custom patch was drawn in the A3921 field, centered on Tucana Dwarf, for ease of analysis. This custom patch covers a field of $\sim 105' \times 70'$. 
The cropped LoVoCCS DECam image of Tucana Dwarf, stacked from $i$, $r$, $g$ band coadded images is shown in Fig. \ref{fig:lovoccs-img}. 

\subsection{Photometric Detection}

The astrometric and photometric measurements for the Hubble images were provided by the Hubble Legacy Archive and generated using the SExtractor software \citep{1996A&AS..117..393B}. The \texttt{lovoccs\_pipe} pipeline generates an object catalog during image processing containing measurements for detected sources. However, the pipeline fails in semi-resolved galaxies like Tucana, where there are diffuse components in addition to resolved stars. New detections and measurements of photometry were generated for the region of the LoVoCCS data containing Tucana Dwarf, using SExtractor. To obtain reliable detections, a composite detection image was created by averaging the $r$, $i$, and $g$-band data to achieve a higher signal-to-noise ratio. Detection and deblending parameters were chosen to prevent blending objects in the dense center of the Tucana Dwarf galaxy, while allowing spurious detections to be removed through color selection and background removal. The detection image was then used as a basis for forced photometry in each of the image bands. We then compared the distribution of photometric magnitudes obtained from the LoVoCCS data to that of the Hubble SExtractor results, validating the measurements of isolated objects detected at high signal-to-noise in both images.

\subsection{Color cut selection}
\begin{figure}[h!]
    \centering
    \includegraphics[width=0.95\linewidth]{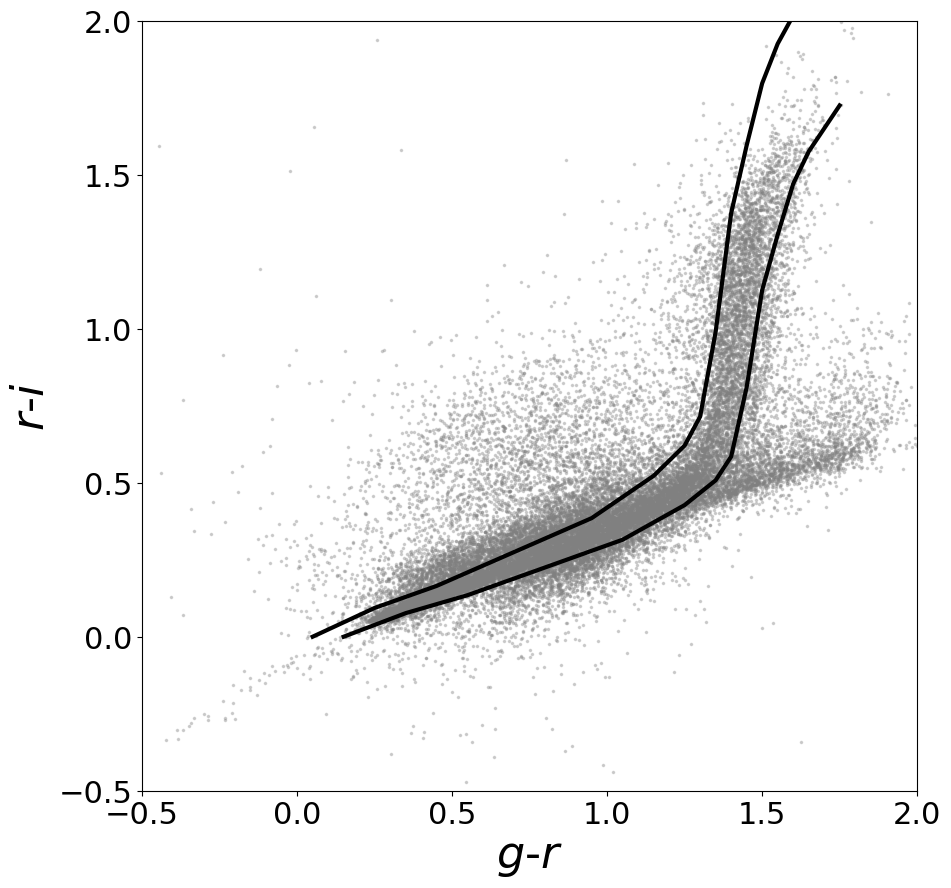}
    \caption{$r$-$i$ vs. $g$-$r$ color-color diagram used to filter for objects consistent with stellar sources in the Tucana Dwarf galaxy. The solid curves show the upper and lower bound criteria for which we made our selection of sources.}
    \label{fig:color-cut}
\end{figure}

\begin{figure}
    \centering
    \includegraphics[width=0.95\linewidth]{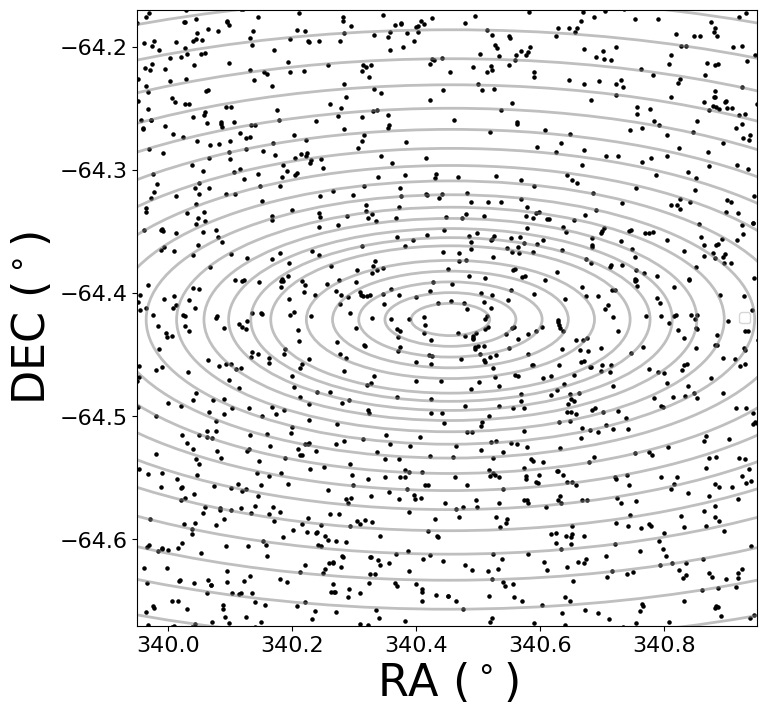}
    \caption{Spatial distribution of objects excluded in color cut selection, outside of first central annulus for which HST data is used. Grey elliptical curves show boundaries for elliptical bins.}
    \label{fig:excluded-objs}
\end{figure}

Based on SExtractor magnitudes of objects within the LoVoCCS catalog, we filtered for objects above 17th magnitude with corresponding errors less than $0.05$. We then constructed a $r$-$i$ vs $g$-$r$ color-color diagram to select for objects in the stellar locus for sources that are consistent with the photometric properties of stellar sources \citep{High_2009}.
The objects in the higher $g$-$r$ end of the color-color diagram that were excluded are potentially high-redshift, compact galaxies \citep{Prakash_2015}.
The objects excluded by the color cut selection criteria were assessed for their spatial distribution. Outside of the central region for which HST imaging is utilized for its resolution, no spatial pattern was found in the excluded sources. The color-color diagram is shown in Fig. \ref{fig:color-cut}. The spatial distribution of objects excluded through the color cut selection is shown in Fig. \ref{fig:excluded-objs}. 

%% file: sections/methods.tex
\section{Processing} \label{sec:methods} 



We performed elliptical fitting over the central region in the HST image to derive the parameters of central coordinates (RA and Dec.), major and minor axes ($a$ and $b$), and tilt angle (relative to the celestial equator). The final values for these structural parameters are shown in Table \ref{table:elliptical-fitting-params}. We found the tilt angle to be close to zero ($0.01_{-0.47 \degree}^{+0.53\degree}$). Subsequently, we divided stellar sources in the HST and LoVoCCS catalogs into concentric elliptical annuli using the derived ellipticity. We then constructed separate stellar density profiles, as functions of semi-major axis distance from the center, for HST and LoVoCCS sources, based on these elliptical bins. To calculate the number density profiles for HST sources, we subtracted the non-imaged area in CCD gaps from the imaging area. Within the wider-field LoVoCCS image, we obtained a background stellar density via iterative linear fitting to the tail-end of the stellar density bins. In each fitting, we treated a number $n$ of the outermost bins as background for linear fitting, and iterated this process over all possible $n$. The linear fit with the minimum slope was chosen as the reference background level for background subtraction. The sensitivity of our final results to this background density will be given further attention in proceeding discussion on bootstrapping. Since the LoVoCCS image does not resolve the center of the galaxy, as in the HST image, the stellar density profiles from the HST and LoVoCCS images were matched in an annulus around the central region and rescaled, in order to combine the two density profiles. We then subtracted the background density from this rescaled and combined density profile. 

Next, we fit a S\'ersic profile over the combined stellar density profile, using Astropy's \verb|Sersic1D| model. In fitting the S\'ersic profile, expressed as $I(R) = I_0 \exp{(-k R^{1/n})}$, the parameters $I_0$, $R$, and $n$ were allowed to vary. The fitted values of $I_0$ and $R$ correspond with the values for the stellar density at the effective radius and the effective (half-light) radius, respectively. The first pass of S\'ersic fitting for both the combined density profile and an HST-only density profile are shown in Fig. \ref{fig:combined_vs_hst_sersic}. The deviation from the S\`ersic fit shown in the preliminary combined profile of Fig. \ref{fig:combined_vs_hst_sersic} may be an artifact of binning, but the S\'ersic fit shows otherwise good agreement with data points. We additionally calculated the change in the S\'ersic fitting with the last five elliptical annuli bins be excluded. With this exclusion of the outermost five bins, the preliminary S\'ersic-fitted half-light radius $r_{eff}$ showed $<1\%$ change in value, indicating a flattened background in the outer bins of the combined stellar number density profile.
\begin{figure*}[h]
    \centering
    \includegraphics[width=1.0\linewidth]{figs/Combined_vs_HST_Sersic_linlog.png}
    \caption{The preliminary S\'ersic profile fitting over stellar density profiles calculated based on (\textit{top}) combined LoVoCCS and HST data, and (\textit{bottom}) HST-only data. The shaded regions indicate area under the curve out to the effective (or, half-light) radii obtained from the two different S\'ersic fits, which are $r_{eff} = 0.0344\degree$ from the combined data and $r_{eff} = 0.0473\degree$ from HST data alone.}
    \label{fig:combined_vs_hst_sersic}
\end{figure*}

\begin{figure*}[h]
    \centering
    \includegraphics[width=0.85\linewidth]{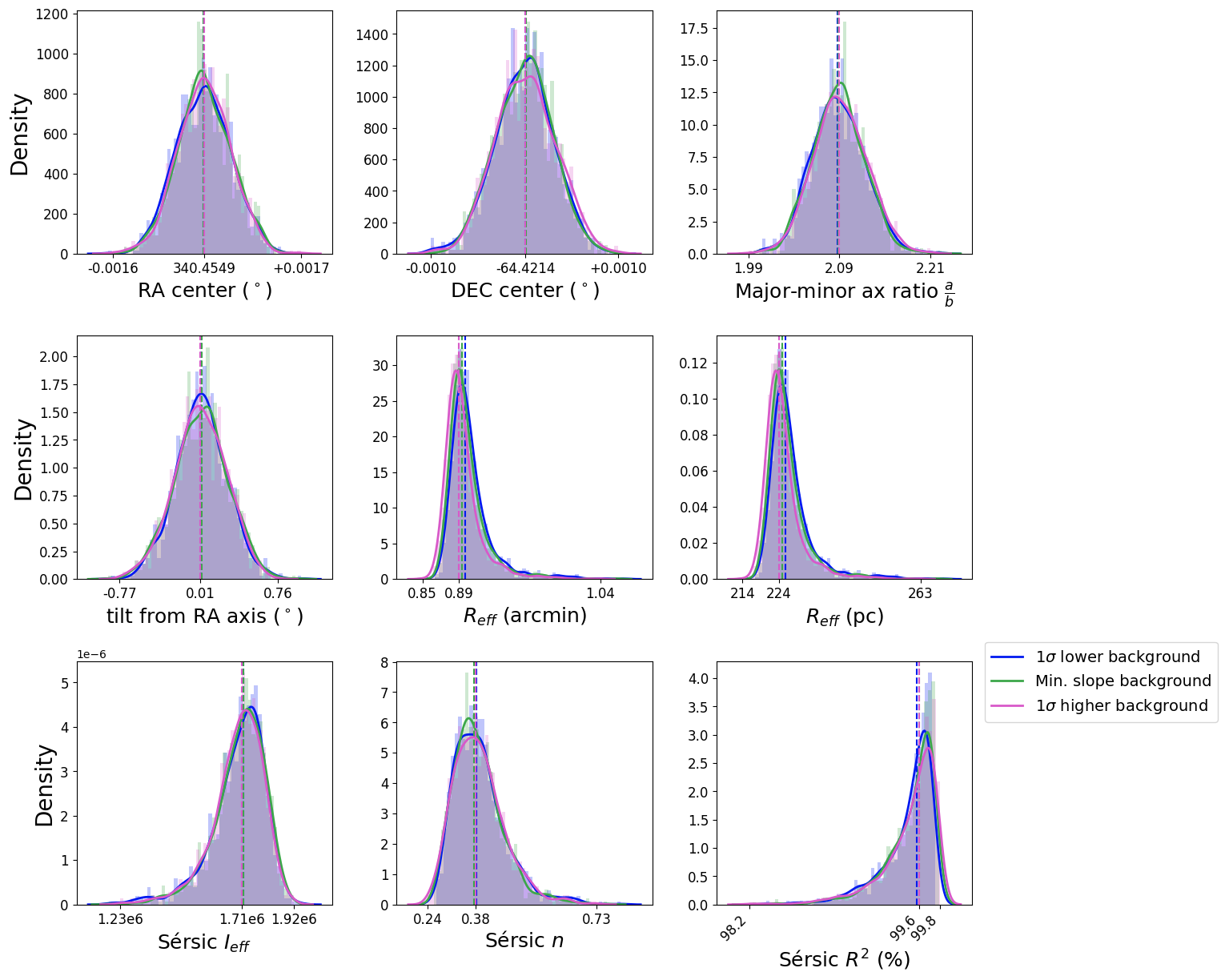}
    \caption{Distributions of fitted parameters across 3,000 total bootstrapping trials, over-plotted with kernel density estimation. Different distributions are plotted for each of the three different background subtraction choices. RA center, DEC center, semi-major axis, semi-minor axis, and tilt angle are the five structural parameters obtained from elliptical fitting over central HST data. Effective (half-light) radius $R_{eff}$, S\'ersic $I_{eff}$, and S\'ersic $n$ are the three parameters obtained from 1D S\'ersic fitting over the stellar density profile. Lastly, the $R^2$ values of S\'ersic fitting over the density profile is presented. 
    } 
    \label{fig:bootstrap-params}
\end{figure*}

After initial S\'ersic fitting to the stellar density profile, we repeated the entire aforementioned processing pipeline through bootstrapping trials to assess the sensitivity of the results to internal instabilities. Bootstrapping trials were ran by resampling all stellar sources without replacement. Then, the steps from color cut selection to S\'ersic fitting were re-run, for a total of 1,000 trials. Furthermore, 1,000 bootstrapping trials were run for 1$\sigma$ lower and 1$\sigma$ higher background subtraction levels, for a total of 3,000 bootstrapping trials, to assess the sensitivity of measured parameters to background subtraction. The distribution of measured parameters obtained across all 3,000 bootstrapping are shown in Fig. \ref{fig:bootstrap-params}.

%% file: sections/results.tex
\section{Results} \label{sec:results} 

We measure the half-light radius to be $R_{e} = 0.89_{-0.02}^{+0.06}$ arcmin (95\% CL), on the semi-major axis, following the definition of the half-light radius in \cite{McConnachie_2012}. The confidence interval was obtained from the distribution of the 1,000 bootstrapping trials carried out with slope-minimizing reference background subtraction. 
Using 870 kpc as the distance to the Tucana dwarf spheroidal galaxy, as measured with the tip of the red giant branch (TRGB) method \citep{saviane_1996}, this angular half-light radius corresponds to a half-light radius of $R_{e} = 225.7_{-5.5}^{+14.4}$ pc.

We compared the preliminary S\'ersic profile fit over combined data from LoVoCCS and HST to the fit over HST-only data, shown in Fig. \ref{fig:combined_vs_hst_sersic}. The HST-only data shows fewer stellar density data points farther from the center of the galaxy, while the wider LoVoCCS imaging supplements data points far from the center. In the combined density profile data points, a flattened extended region is clearly visible. 

Given the elliptical fitting of the central HST image used to create elliptical bins and construct the stellar density profiles, the structural parameters for Tucana Dwarf from this fitting are presented in Table \ref{table:elliptical-fitting-params}. We calculated the uncertainties of these parameters from the bootstrapping process, with the distribution of structural parameters from elliptical fitting approximated as symmetric. 

\begin{table}[h!]
    \centering
    \begin{tabular}{|l|c|c|} 
        \hline
         Parameter &  Value & 95\% CL \\
         \hline
         Center RA (\degree) & 340.45 & 0.00087 \\
         Center DEC (\degree) & -64.42 & 0.00062 \\
         Major-to-minor axes $\frac{a}{b}$ & 2.096 & 0.062 \\
         Tilt from RA axis (\degree) & 0.01 & [-0.47, +0.53] \\
         \hline
         S\'ersic $R_{eff}$ (arcmin) & 0.89 & [-0.02, +0.06]  \\
         S\'ersic $I_{eff}$ & $1.72\times10^6$ & [$-2.62,+1.44$]$\times10^5$ \\
         S\'ersic $n$ & 0.38 & [-0.10, +0.18] \\
         S\'ersic fit $R^2$ & 0.996 & [-0.006, +0.001] \\
         \hline
    \end{tabular}
    \caption{
    Distribution of elliptical fitting and S\'ersic fitting parameters. Uncertainties are calculated from bootstrapping. A single value for uncertainty is used when the distribution is roughly symmetric.
    \label{table:elliptical-fitting-params}
    }
\end{table}

The results from the bootstrapping process are shown in Fig. \ref{fig:bootstrap-params}. The structural parameters derived from elliptical fitting in the boostrapping (RA and DEC for the centroid, the major-to-minor axes ratio, and the tilt angle from RA axis) show roughly symmetric distributions with small uncertainties.

%% file: sections/discussion.tex
\section{Discussion} \label{sec:discussion}
Over the past three decades, a redshift-independent distance to Tucana has been measured to lie between 870 and 916 kpc using the TRGB method \citep{lavery_1991, castellani1995tucanadwarfgalaxy, saviane_1996, Jacobs_2009, Tully_2013}, and $890 \pm 50$ kpc with the RR Lyrae method \citep{Bernard_2009}. Therefore, uncertainties in the distance to Tucana Dwarf affect the physical half-light radius measure. Our measurement of the half-light radius is better constrained in terms of angle subtended, at $0.89_{-0.02}^{+0.06}$ arcmin. 

The previous measurement of the half-light radius of Tucana Dwarf by \cite{saviane_1996} was based on imaging from EFOSC2 at the ESO 2.2m telescope, with a field-of-view of $5.7' \times 5.7'$. Compared to this previous work, deeper and wider imaging were available to us through the combination of HST and LoVoCCS data. 
The half-light radius, referred to as the `effective geometric radius' in \cite{saviane_1996}, was derived from an exponential fit of the brightness profile based on a smoothed image, compared to our analysis utilizing resolved stellar populations. The previous results measure an effective radius, or half-light radius, of $R_e = 205 \pm 32$ pc, from the exponential fit. 

From the comparison of combined LoVoCCS and HST versus HST-only stellar density profiles in Fig. \ref{fig:combined_vs_hst_sersic}, it is apparent that with HST data alone, it is not possible to observe the flattening of the density profile to its background value within the field of the HST image. The wider imaging from LoVoCCS has allowed for measurement of stellar number density out to a much larger radius, providing a clearly flattened profile far from the center. The combination of high-resolution central HST data and extended LoVoCCS data presents significantly improved constraining power for the half-light radius of the Tucana Dwarf galaxy. 

The bootstrapping process, performed with varying levels of background subtraction, shows both internal consistency of our results based on our data, as well as a justification for the background subtraction choice, to which the final results have no significant sensitivity. Of the structural elliptical fitting parameters shown in Table \ref{table:elliptical-fitting-params}, the low uncertainties show consistent results, with the tilt angle relative to the RA axis being small enough ($0.01_{-0.47  \degree}^{+0.53  \degree}$) to have insignificant effects on the final measurements. Of the values derived from S\'ersic fitting shown in Table \ref{table:elliptical-fitting-params}, the large uncertainties in the S\'ersic $I_{eff}$ parameter reflects the high sensitivity of S\'ersic profile fitting to $I_{eff}$. 

Based on recent kinematic studies of Tucana Dwarf \citep{Fraternali_2009, Gregory_2019, Taibi_2020}, there are few member objects with velocity measurements within the galaxy. Future observations would be necessary to help constrain the mass of the galaxy, in conjunction with the density profiles obtained in this work. Furthermore, kinematic studies of isolated dwarf galaxies such as Tucana would help disentangle the kinematic effects of dark matter and resonance processes between the orbital period of a dwarf and its natural radial oscillation period \citep{Mateo_1998}. We believe that with our updated measurement of the half-light radius of the Tucana Dwarf galaxy, the uncertainty of its mass estimate now lies within its kinematics, highlighting the need for a spectroscopic follow-up on Tucana Dwarf.

%% file: ref.bib
@ARTICLE{2022ApJ...935..167A,
       author = {{Astropy Collaboration} and {Price-Whelan}, Adrian M. and {Lim}, Pey Lian and {Earl}, Nicholas and {Starkman}, Nathaniel and {Bradley}, Larry and {Shupe}, David L. and {Patil}, Aarya A. and {Corrales}, Lia and {Brasseur}, C.~E. and {N{\"o}the}, Maximilian and {Donath}, Axel and {Tollerud}, Erik and {Morris}, Brett M. and {Ginsburg}, Adam and {Vaher}, Eero and {Weaver}, Benjamin A. and {Tocknell}, James and {Jamieson}, William and {van Kerkwijk}, Marten H. and {Robitaille}, Thomas P. and {Merry}, Bruce and {Bachetti}, Matteo and {G{\"u}nther}, H. Moritz and {Aldcroft}, Thomas L. and {Alvarado-Montes}, Jaime A. and {Archibald}, Anne M. and {B{\'o}di}, Attila and {Bapat}, Shreyas and {Barentsen}, Geert and {Baz{\'a}n}, Juanjo and {Biswas}, Manish and {Boquien}, M{\'e}d{\'e}ric and {Burke}, D.~J. and {Cara}, Daria and {Cara}, Mihai and {Conroy}, Kyle E. and {Conseil}, Simon and {Craig}, Matthew W. and {Cross}, Robert M. and {Cruz}, Kelle L. and {D'Eugenio}, Francesco and {Dencheva}, Nadia and {Devillepoix}, Hadrien A.~R. and {Dietrich}, J{\"o}rg P. and {Eigenbrot}, Arthur Davis and {Erben}, Thomas and {Ferreira}, Leonardo and {Foreman-Mackey}, Daniel and {Fox}, Ryan and {Freij}, Nabil and {Garg}, Suyog and {Geda}, Robel and {Glattly}, Lauren and {Gondhalekar}, Yash and {Gordon}, Karl D. and {Grant}, David and {Greenfield}, Perry and {Groener}, Austen M. and {Guest}, Steve and {Gurovich}, Sebastian and {Handberg}, Rasmus and {Hart}, Akeem and {Hatfield-Dodds}, Zac and {Homeier}, Derek and {Hosseinzadeh}, Griffin and {Jenness}, Tim and {Jones}, Craig K. and {Joseph}, Prajwel and {Kalmbach}, J. Bryce and {Karamehmetoglu}, Emir and {Ka{\l}uszy{\'n}ski}, Miko{\l}aj and {Kelley}, Michael S.~P. and {Kern}, Nicholas and {Kerzendorf}, Wolfgang E. and {Koch}, Eric W. and {Kulumani}, Shankar and {Lee}, Antony and {Ly}, Chun and {Ma}, Zhiyuan and {MacBride}, Conor and {Maljaars}, Jakob M. and {Muna}, Demitri and {Murphy}, N.~A. and {Norman}, Henrik and {O'Steen}, Richard and {Oman}, Kyle A. and {Pacifici}, Camilla and {Pascual}, Sergio and {Pascual-Granado}, J. and {Patil}, Rohit R. and {Perren}, Gabriel I. and {Pickering}, Timothy E. and {Rastogi}, Tanuj and {Roulston}, Benjamin R. and {Ryan}, Daniel F. and {Rykoff}, Eli S. and {Sabater}, Jose and {Sakurikar}, Parikshit and {Salgado}, Jes{\'u}s and {Sanghi}, Aniket and {Saunders}, Nicholas and {Savchenko}, Volodymyr and {Schwardt}, Ludwig and {Seifert-Eckert}, Michael and {Shih}, Albert Y. and {Jain}, Anany Shrey and {Shukla}, Gyanendra and {Sick}, Jonathan and {Simpson}, Chris and {Singanamalla}, Sudheesh and {Singer}, Leo P. and {Singhal}, Jaladh and {Sinha}, Manodeep and {Sip{\H{o}}cz}, Brigitta M. and {Spitler}, Lee R. and {Stansby}, David and {Streicher}, Ole and {{\v{S}}umak}, Jani and {Swinbank}, John D. and {Taranu}, Dan S. and {Tewary}, Nikita and {Tremblay}, Grant R. and {de Val-Borro}, Miguel and {Van Kooten}, Samuel J. and {Vasovi{\'c}}, Zlatan and {Verma}, Shresth and {de Miranda Cardoso}, Jos{\'e} Vin{\'\i}cius and {Williams}, Peter K.~G. and {Wilson}, Tom J. and {Winkel}, Benjamin and {Wood-Vasey}, W.~M. and {Xue}, Rui and {Yoachim}, Peter and {Zhang}, Chen and {Zonca}, Andrea and {Astropy Project Contributors}},
        title = "{The Astropy Project: Sustaining and Growing a Community-oriented Open-source Project and the Latest Major Release (v5.0) of the Core Package}",
      journal = {\apj},
         year = 2022,
        month = aug,
       volume = {935},
       number = {2},
          eid = {167},
        pages = {167},
          doi = {10.3847/1538-4357/ac7c74},
archivePrefix = {arXiv},
       eprint = {2206.14220},
 primaryClass = {astro-ph.IM},
       adsurl = {https://ui.adsabs.harvard.edu/abs/2022ApJ...935..167A}
}

@ARTICLE{2018AJ....156..123A,
       author = {{Astropy Collaboration} and {Price-Whelan}, A.~M. and {Sip{\H{o}}cz}, B.~M. and {G{\"u}nther}, H.~M. and {Lim}, P.~L. and {Crawford}, S.~M. and {Conseil}, S. and {Shupe}, D.~L. and {Craig}, M.~W. and {Dencheva}, N. and {Ginsburg}, A. and {VanderPlas}, J.~T. and {Bradley}, L.~D. and {P{\'e}rez-Su{\'a}rez}, D. and {de Val-Borro}, M. and {Aldcroft}, T.~L. and {Cruz}, K.~L. and {Robitaille}, T.~P. and {Tollerud}, E.~J. and {Ardelean}, C. and {Babej}, T. and {Bach}, Y.~P. and {Bachetti}, M. and {Bakanov}, A.~V. and {Bamford}, S.~P. and {Barentsen}, G. and {Barmby}, P. and {Baumbach}, A. and {Berry}, K.~L. and {Biscani}, F. and {Boquien}, M. and {Bostroem}, K.~A. and {Bouma}, L.~G. and {Brammer}, G.~B. and {Bray}, E.~M. and {Breytenbach}, H. and {Buddelmeijer}, H. and {Burke}, D.~J. and {Calderone}, G. and {Cano Rodr{\'\i}guez}, J.~L. and {Cara}, M. and {Cardoso}, J.~V.~M. and {Cheedella}, S. and {Copin}, Y. and {Corrales}, L. and {Crichton}, D. and {D'Avella}, D. and {Deil}, C. and {Depagne}, {\'E}. and {Dietrich}, J.~P. and {Donath}, A. and {Droettboom}, M. and {Earl}, N. and {Erben}, T. and {Fabbro}, S. and {Ferreira}, L.~A. and {Finethy}, T. and {Fox}, R.~T. and {Garrison}, L.~H. and {Gibbons}, S.~L.~J. and {Goldstein}, D.~A. and {Gommers}, R. and {Greco}, J.~P. and {Greenfield}, P. and {Groener}, A.~M. and {Grollier}, F. and {Hagen}, A. and {Hirst}, P. and {Homeier}, D. and {Horton}, A.~J. and {Hosseinzadeh}, G. and {Hu}, L. and {Hunkeler}, J.~S. and {Ivezi{\'c}}, {\v{Z}}. and {Jain}, A. and {Jenness}, T. and {Kanarek}, G. and {Kendrew}, S. and {Kern}, N.~S. and {Kerzendorf}, W.~E. and {Khvalko}, A. and {King}, J. and {Kirkby}, D. and {Kulkarni}, A.~M. and {Kumar}, A. and {Lee}, A. and {Lenz}, D. and {Littlefair}, S.~P. and {Ma}, Z. and {Macleod}, D.~M. and {Mastropietro}, M. and {McCully}, C. and {Montagnac}, S. and {Morris}, B.~M. and {Mueller}, M. and {Mumford}, S.~J. and {Muna}, D. and {Murphy}, N.~A. and {Nelson}, S. and {Nguyen}, G.~H. and {Ninan}, J.~P. and {N{\"o}the}, M. and {Ogaz}, S. and {Oh}, S. and {Parejko}, J.~K. and {Parley}, N. and {Pascual}, S. and {Patil}, R. and {Patil}, A.~A. and {Plunkett}, A.~L. and {Prochaska}, J.~X. and {Rastogi}, T. and {Reddy Janga}, V. and {Sabater}, J. and {Sakurikar}, P. and {Seifert}, M. and {Sherbert}, L.~E. and {Sherwood-Taylor}, H. and {Shih}, A.~Y. and {Sick}, J. and {Silbiger}, M.~T. and {Singanamalla}, S. and {Singer}, L.~P. and {Sladen}, P.~H. and {Sooley}, K.~A. and {Sornarajah}, S. and {Streicher}, O. and {Teuben}, P. and {Thomas}, S.~W. and {Tremblay}, G.~R. and {Turner}, J.~E.~H. and {Terr{\'o}n}, V. and {van Kerkwijk}, M.~H. and {de la Vega}, A. and {Watkins}, L.~L. and {Weaver}, B.~A. and {Whitmore}, J.~B. and {Woillez}, J. and {Zabalza}, V. and {Astropy Contributors}},
        title = "{The Astropy Project: Building an Open-science Project and Status of the v2.0 Core Package}",
      journal = {\aj},
         year = 2018,
        month = sep,
       volume = {156},
       number = {3},
          eid = {123},
        pages = {123},
          doi = {10.3847/1538-3881/aabc4f},
archivePrefix = {arXiv},
       eprint = {1801.02634},
 primaryClass = {astro-ph.IM},
       adsurl = {https://ui.adsabs.harvard.edu/abs/2018AJ....156..123A}
}

@article{2013A&A...558A..33A,
       author = {{Astropy Collaboration} and {Robitaille}, Thomas P. and
         {Tollerud}, Erik J. and {Greenfield}, Perry and {Droettboom}, Michael and
         {Bray}, Erik and {Aldcroft}, Tom and {Davis}, Matt and
         {Ginsburg}, Adam and {Price-Whelan}, Adrian M. and
         {Kerzendorf}, Wolfgang E. and {Conley}, Alexander and {Crighton}, Neil and
         {Barbary}, Kyle and {Muna}, Demitri and {Ferguson}, Henry and
         {Grollier}, Fr{\'e}d{\'e}ric and {Parikh}, Madhura M. and
         {Nair}, Prasanth H. and {Unther}, Hans M. and {Deil}, Christoph and
         {Woillez}, Julien and {Conseil}, Simon and {Kramer}, Roban and
         {Turner}, James E.~H. and {Singer}, Leo and {Fox}, Ryan and
         {Weaver}, Benjamin A. and {Zabalza}, Victor and {Edwards}, Zachary I. and
         {Azalee Bostroem}, K. and {Burke}, D.~J. and {Casey}, Andrew R. and
         {Crawford}, Steven M. and {Dencheva}, Nadia and {Ely}, Justin and
         {Jenness}, Tim and {Labrie}, Kathleen and {Lim}, Pey Lian and
         {Pierfederici}, Francesco and {Pontzen}, Andrew and {Ptak}, Andy and
         {Refsdal}, Brian and {Servillat}, Mathieu and {Streicher}, Ole},
        title = "{Astropy: A community Python package for astronomy}",
      journal = {\aap},
         year = "2013",
        month = "Oct",
       volume = {558},
          eid = {A33},
        pages = {A33},
          doi = {10.1051/0004-6361/201322068},
archivePrefix = {arXiv},
       eprint = {1307.6212},
 primaryClass = {astro-ph.IM},
       adsurl = {https://ui.adsabs.harvard.edu/abs/2013A&A...558A..33A}
}

@ARTICLE{1996A&AS..117..393B,
       author = {{Bertin}, E. and {Arnouts}, S.},
        title = "{SExtractor: Software for source extraction.}",
      journal = {\aaps},
         year = "1996",
        month = "Jun",
       volume = {117},
        pages = {393-404},
          doi = {10.1051/aas:1996164},
       adsurl = {https://ui.adsabs.harvard.edu/abs/1996A&AS..117..393B}
}

@article{battaglia2022stellardynamicsdarkmatter,
    title={Stellar dynamics and dark matter in Local Group dwarf galaxies}, 
    author={Giuseppina Battaglia and Carlo Nipoti},
    year={2022},
    journal={Nature Astronomy},
    vol={6},
    number={659},
    url={https://arxiv.org/abs/2205.07821}, 
}

@misc{saviane_1996,
      title={CCD photometry of the Tucana dwarf galaxy}, 
      author={Ivo Saviane and Enrico V. Held and Giampaolo Piotto},
      year={1996},
      eprint={astro-ph/9601165},
      archivePrefix={arXiv},
      primaryClass={astro-ph},
      url={https://arxiv.org/abs/astro-ph/9601165}, 
}

@article{Fraternali_2009,
   title={Life at the periphery of the Local Group: the kinematics of the Tucana dwarf galaxy},
   volume={499},
   ISSN={1432-0746},
   url={http://dx.doi.org/10.1051/0004-6361/200810830},
   DOI={10.1051/0004-6361/200810830},
   number={1},
   journal={Astronomy \& Astrophysics},
   publisher={EDP Sciences},
   author={Fraternali, F. and Tolstoy, E. and Irwin, M. J. and Cole, A. A.},
   year={2009},
   month=apr, pages={121–128} }

@article{Gregory_2019,
   title={Kinematics of the Tucana Dwarf Galaxy: an unusually dense dwarf in the Local Group},
   volume={485},
   ISSN={1365-2966},
   url={http://dx.doi.org/10.1093/mnras/stz518},
   DOI={10.1093/mnras/stz518},
   number={2},
   journal={Monthly Notices of the Royal Astronomical Society},
   publisher={Oxford University Press (OUP)},
   author={Gregory, Alexandra L and Collins, Michelle L M and Read, Justin I and Irwin, Michael J and Ibata, Rodrigo A and Martin, Nicolas F and McConnachie, Alan W and Weisz, Daniel R},
   year={2019},
   month=feb, pages={2010–2025} }

@article{Taibi_2020,
   title={The Tucana dwarf spheroidal galaxy: not such a massive failure after all},
   volume={635},
   ISSN={1432-0746},
   url={http://dx.doi.org/10.1051/0004-6361/201937240},
   DOI={10.1051/0004-6361/201937240},
   journal={Astronomy \& Astrophysics},
   publisher={EDP Sciences},
   author={Taibi, S. and Battaglia, G. and Rejkuba, M. and Leaman, R. and Kacharov, N. and Iorio, G. and Jablonka, P. and Zoccali, M.},
   year={2020},
   month=mar, pages={A152} }

@article{flaugher_dark_2015,
	title = {{THE} {DARK} {ENERGY} {CAMERA}},
	volume = {150},
	issn = {1538-3881},
	url = {https://dx.doi.org/10.1088/0004-6256/150/5/150},
	doi = {10.1088/0004-6256/1504/5/150},
	language = {en},
	number = {5},
	urldate = {2025-09-24},
	journal = {The Astronomical Journal},
	author = {Flaugher, B. and Diehl, H. T. and Honscheid, K. and Abbott, T. M. C. and Alvarez, O. and Angstadt, R. and Annis, J. T. and Antonik, M. and Ballester, O. and Beaufore, L. and Bernstein, G. M. and Bernstein, R. A. and Bigelow, B. and Bonati, M. and Boprie, D. and Brooks, D. and Buckley-Geer, E. J. and Campa, J. and Cardiel-Sas, L. and Castander, F. J. and Castilla, J. and Cease, H. and Cela-Ruiz, J. M. and Chappa, S. and Chi, E. and Cooper, C. and da Costa, L. N. and Dede, E. and Derylo, G. and DePoy, D. L. and de Vicente, J. and Doel, P. and Drlica-Wagner, A. and Eiting, J. and Elliott, A. E. and Emes, J. and Estrada, J. and Fausti Neto, A. and Finley, D. A. and Flores, R. and Frieman, J. and Gerdes, D. and Gladders, M. D. and Gregory, B. and Gutierrez, G. R. and Hao, J. and Holland, S. E. and Holm, S. and Huffman, D. and Jackson, C. and James, D. J. and Jonas, M. and Karcher, A. and Karliner, I. and Kent, S. and Kessler, R. and Kozlovsky, M. and Kron, R. G. and Kubik, D. and Kuehn, K. and Kuhlmann, S. and Kuk, K. and Lahav, O. and Lathrop, A. and Lee, J. and Levi, M. E. and Lewis, P. and Li, T. S. and Mandrichenko, I. and Marshall, J. L. and Martinez, G. and Merritt, K. W. and Miquel, R. and Muñoz, F. and Neilsen, E. H. and Nichol, R. C. and Nord, B. and Ogando, R. and Olsen, J. and Palaio, N. and Patton, K. and Peoples, J. and Plazas, A. A. and Rauch, J. and Reil, K. and Rheault, J.-P. and Roe, N. A. and Rogers, H. and Roodman, A. and Sanchez, E. and Scarpine, V. and Schindler, R. H. and Schmidt, R. and Schmitt, R. and Schubnell, M. and Schultz, K. and Schurter, P. and Scott, L. and Serrano, S. and Shaw, T. M. and Smith, R. C. and Soares-Santos, M. and Stefanik, A. and Stuermer, W. and Suchyta, E. and Sypniewski, A. and Tarle, G. and Thaler, J. and Tighe, R. and Tran, C. and Tucker, D. and Walker, A. R. and Wang, G. and Watson, M. and Weaverdyck, C. and Wester, W. and Woods, R. and Yanny, B. and Collaboration), (The DES},
	month = oct,
	year = {2015},
	note = {Publisher: The American Astronomical Society},
	pages = {150},
}

@article{fu_lovoccs_2022,
	title = {{LoVoCCS}. {I}. {Survey} {Introduction}, {Data} {Processing} {Pipeline}, and {Early} {Science} {Results}},
	volume = {933},
	issn = {0004-637X},
	url = {https://ui.adsabs.harvard.edu/abs/2022ApJ...933...84F},
	doi = {10.3847/1538-4357/ac68e8},
	urldate = {2025-09-24},
	journal = {The Astrophysical Journal},
	author = {Fu, Shenming and Dell'Antonio, Ian and Chary, Ranga-Ram and Clowe, Douglas and Cooper, M. C. and Donahue, Megan and Evrard, August and Lacy, Mark and Lauer, Tod and Liu, Binyang and McCleary, Jacqueline and Meneghetti, Massimo and Miyatake, Hironao and Montes, Mireia and Natarajan, Priyamvada and Ntampaka, Michelle and Pierpaoli, Elena and Postman, Marc and Sohn, Jubee and Umetsu, Keiichi and Utsumi, Yousuke and Wilson, Gillian},
	month = jul,
	year = {2022},
	note = {ADS Bibcode: 2022ApJ...933...84F},
	pages = {84},
}

@INPROCEEDINGS{englert_aas_2025,
       author = {{Englert}, Anthony and {Dell'Antonio}, Ian and {Fu}, Shenming and {Escalante}, Zacharias and {Helhoski}, Soren and {Nelson}, Jessica},
        title = "{Recent Results from the Local Volume Complete Cluster Survey}",
    booktitle = {American Astronomical Society Meeting Abstracts \#245},
         year = 2025,
       series = {American Astronomical Society Meeting Abstracts},
       volume = {245},
        month = jan,
          eid = {412.07},
        pages = {412.07},
       adsurl = {https://ui.adsabs.harvard.edu/abs/2025AAS...24541207E}
}

@article{englert_intracluster_2025,
	title = {The {Intracluster} {Light} of {Abell} 3667: {Unveiling} an {Optical} {Bridge} in {LSST} {Precursor} {Data}},
	volume = {989},
	issn = {2041-8205},
	shorttitle = {The {Intracluster} {Light} of {Abell} 3667},
	url = {https://dx.doi.org/10.3847/2041-8213/ade8f1},
	doi = {10.3847/2041-8213/ade8f1},
	language = {en},
	number = {1},
	urldate = {2025-09-24},
	journal = {The Astrophysical Journal Letters},
	author = {Englert, Anthony and Dell'Antonio, Ian and Montes, Mireia},
	month = aug,
	year = {2025},
	note = {Publisher: The American Astronomical Society},
	pages = {L2},
}

@misc{bosch2018overviewlsstimageprocessing,
      title={An Overview of the LSST Image Processing Pipelines}, 
      author={James Bosch and Yusra AlSayyad and Robert Armstrong and Eric Bellm and Hsin-Fang Chiang and Siegfried Eggl and Krzysztof Findeisen and Merlin Fisher-Levine and Leanne P. Guy and Augustin Guyonnet and Željko Ivezić and Tim Jenness and Gábor Kovács and K. Simon Krughoff and Robert H. Lupton and Nate B. Lust and Lauren A. MacArthur and Joshua Meyers and Fred Moolekamp and Christopher B. Morrison and Timothy D. Morton and William O'Mullane and John K. Parejko and Andrés A. Plazas and Paul A. Price and Meredith L. Rawls and Sophie L. Reed and Pim Schellart and Colin T. Slater and Ian Sullivan and John. D. Swinbank and Dan Taranu and Christopher Z. Waters and W. M. Wood-Vasey},
      year={2018},
      eprint={1812.03248},
      archivePrefix={arXiv},
      primaryClass={astro-ph.IM},
      url={https://arxiv.org/abs/1812.03248}, 
}

@misc{bosch_overview_2019,
	address = {eprint: arXiv:1812.03248},
	title = {An {Overview} of the {LSST} {Image} {Processing} {Pipelines}},
	url = {https://ui.adsabs.harvard.edu/abs/2019ASPC..523..521B},
	doi = {10.48550/arXiv.1812.03248},
	urldate = {2025-09-24},
	publisher = {arXiv},
	author = {Bosch, James and AlSayyad, Yusra and Armstrong, Robert and Bellm, Eric and Chiang, Hsin-Fang and Eggl, Siegfried and Findeisen, Krzysztof and Fisher-Levine, Merlin and Guy, Leanne P. and Guyonnet, Augustin and Ivezić, Željko and Jenness, Tim and Kovács, Gábor and Krughoff, K. Simon and Lupton, Robert H. and Lust, Nate B. and MacArthur, Lauren A. and Meyers, Joshua and Moolekamp, Fred and Morrison, Christopher B. and Morton, Timothy D. and O'Mullane, William and Parejko, John K. and Plazas, Andrés A. and Price, Paul A. and Rawls, Meredith L. and Reed, Sophie L. and Schellart, Pim and Slater, Colin T. and Sullivan, Ian and Swinbank, John D. and Taranu, Dan and Waters, Christopher Z. and Wood-Vasey, W. M.},
	month = oct,
	year = {2019},
	note = {Conference Name: Astronomical Data Analysis Software and Systems XXVII
Volume: 523
ADS Bibcode: 2019ASPC..523..521B},
}

@article{Savino_2019,
   title={The HST/ACS star formation history of the Tucana dwarf spheroidal galaxy: clues from the horizontal branch},
   volume={630},
   ISSN={1432-0746},
   url={http://dx.doi.org/10.1051/0004-6361/201936077},
   DOI={10.1051/0004-6361/201936077},
   journal={Astronomy \& Astrophysics},
   publisher={EDP Sciences},
   author={Savino, A. and Tolstoy, E. and Salaris, M. and Monelli, M. and de Boer, T. J. L.},
   year={2019},
   month=oct, pages={A116} }

@article{Monelli_2010,
   title={THE ACS LCID PROJECT. VI. THE STAR FORMATION HISTORY OF THE TUCANA dSph AND THE RELATIVE AGES OF THE ISOLATED dSph GALAXIES},
   volume={722},
   ISSN={1538-4357},
   url={http://dx.doi.org/10.1088/0004-637X/722/2/1864},
   DOI={10.1088/0004-637x/722/2/1864},
   number={2},
   journal={The Astrophysical Journal},
   publisher={American Astronomical Society},
   author={Monelli, M. and Gallart, C. and Hidalgo, S. L. and Aparicio, A. and Skillman, E. D. and Cole, A. A. and Weisz, D. R. and Mayer, L. and Bernard, E. J. and Cassisi, S. and Dolphin, A. E. and Drozdovsky, I. and Stetson, P. B.},
   year={2010},
   month=oct, pages={1864–1878} }

@article{Teyssier_2012,
    author = {Teyssier, Maureen and Johnston, Kathryn V. and Kuhlen, Michael},
    title = {Identifying Local Group field galaxies that have interacted with the Milky Way},
    journal = {Monthly Notices of the Royal Astronomical Society},
    volume = {426},
    number = {3},
    pages = {1808-1818},
    year = {2012},
    month = {11},
    issn = {0035-8711},
    doi = {10.1111/j.1365-2966.2012.21793.x},
    url = {https://doi.org/10.1111/j.1365-2966.2012.21793.x},
    eprint = {https://academic.oup.com/mnras/article-pdf/426/3/1808/3100886/426-3-1808.pdf},
}

@article{Sales_2007,
   title={Cosmic menage a trois: the origin of satellite galaxies on extreme orbits},
   volume={379},
   ISSN={1365-2966},
   url={http://dx.doi.org/10.1111/j.1365-2966.2007.12026.x},
   DOI={10.1111/j.1365-2966.2007.12026.x},
   number={4},
   journal={Monthly Notices of the Royal Astronomical Society},
   publisher={Oxford University Press (OUP)},
   author={Sales, L. V. and Navarro, J. F. and Abadi, M. G. and Steinmetz, M.},
   year={2007},
   month=aug, pages={1475–1483} }

@article{Santos_Santos_2023,
   title={The Tucana dwarf spheroidal: a distant backsplash galaxy of M31?},
   volume={520},
   ISSN={1365-2966},
   url={http://dx.doi.org/10.1093/mnras/stad085},
   DOI={10.1093/mnras/stad085},
   number={1},
   journal={Monthly Notices of the Royal Astronomical Society},
   publisher={Oxford University Press (OUP)},
   author={Santos-Santos, Isabel M E and Navarro, Julio F and McConnachie, Alan},
   year={2023},
   month=jan, pages={55–62} }

@article{McConnachie_2012,
   title={THE OBSERVED PROPERTIES OF DWARF GALAXIES IN AND AROUND THE LOCAL GROUP},
   volume={144},
   ISSN={1538-3881},
   url={http://dx.doi.org/10.1088/0004-6256/144/1/4},
   DOI={10.1088/0004-6256/144/1/4},
   number={1},
   journal={The Astronomical Journal},
   publisher={American Astronomical Society},
   author={McConnachie, Alan W.},
   year={2012},
   month=jun, pages={4} }

@misc{castellani1995tucanadwarfgalaxy,
      title={The {T}ucana dwarf galaxy}, 
      author={Marco Castellani and Gianni Marconi and Roberto Buonanno},
      year={1995},
      eprint={astro-ph/9511144},
      archivePrefix={arXiv},
      primaryClass={astro-ph},
      url={https://arxiv.org/abs/astro-ph/9511144}, 
}

@article{lavery_1991,
    title={A New Member of the Local Group: The {T}ucana Dwarf Galaxy
},
    author={Lavery, Russell J. and Mighell, Kenneth J.},
    year={1992},
    month=jan,
    pages={3},
    volume={103},
    url={https://legacy.adsabs.harvard.edu/full/1992AJ....103...81L},
    DOI={10.1086/116042},
    number={1},
    journal={The Astronomical Journal},
   publisher={American Astronomical Society},
}

@article{Bernard_2009,
   title={THE ACS LCID PROJECT. I. SHORT-PERIOD VARIABLES IN THE ISOLATED DWARF SPHEROIDAL GALAXIES CETUS AND TUCANA},
   volume={699},
   ISSN={1538-4357},
   url={http://dx.doi.org/10.1088/0004-637X/699/2/1742},
   DOI={10.1088/0004-637x/699/2/1742},
   number={2},
   journal={The Astrophysical Journal},
   publisher={American Astronomical Society},
   author={Bernard, Edouard J. and Monelli, Matteo and Gallart, Carme and Drozdovsky, Igor and Stetson, Peter B. and Aparicio, Antonio and Cassisi, Santi and Mayer, Lucio and Cole, Andrew A. and Hidalgo, Sebastian L. and Skillman, Evan D. and Tolstoy, Eline},
   year={2009},
   month=jun, pages={1742–1764} }

@article{Jacobs_2009,
   title={THE EXTRAGALACTIC DISTANCE DATABASE: COLOR-MAGNITUDE DIAGRAMS},
   volume={138},
   ISSN={1538-3881},
   url={http://dx.doi.org/10.1088/0004-6256/138/2/332},
   DOI={10.1088/0004-6256/138/2/332},
   number={2},
   journal={The Astronomical Journal},
   publisher={American Astronomical Society},
   author={Jacobs, Bradley A. and Rizzi, Luca and Tully, R. Brent and Shaya, Edward J. and Makarov, Dmitry I. and Makarova, Lidia},
   year={2009},
   month=jun, pages={332–337} }

@article{Tully_2013,
   title={COSMICFLOWS-2: THE DATA},
   volume={146},
   ISSN={1538-3881},
   url={http://dx.doi.org/10.1088/0004-6256/146/4/86},
   DOI={10.1088/0004-6256/146/4/86},
   number={4},
   journal={The Astronomical Journal},
   publisher={American Astronomical Society},
   author={Tully, R. Brent and Courtois, Hélène M. and Dolphin, Andrew E. and Fisher, J. Richard and Héraudeau, Philippe and Jacobs, Bradley A. and Karachentsev, Igor D. and Makarov, Dmitry and Makarova, Lidia and Mitronova, Sofia and Rizzi, Luca and Shaya, Edward J. and Sorce, Jenny G. and Wu, Po-Feng},
   year={2013},
   month=sep, pages={86} }

@article{Bernard_2008,
   title={The ACS LCID Project: RR Lyrae Stars as Tracers of Old Population Gradients in the Isolated Dwarf Spheroidal Galaxy Tucana},
   volume={678},
   ISSN={1538-4357},
   url={http://dx.doi.org/10.1086/588285},
   DOI={10.1086/588285},
   number={1},
   journal={The Astrophysical Journal},
   publisher={American Astronomical Society},
   author={Bernard, Edouard J. and Gallart, Carme and Monelli, Matteo and Aparicio, Antonio and Cassisi, Santi and Skillman, Evan D. and Stetson, Peter B. and Cole, Andrew A. and Drozdovsky, Igor and Hidalgo, Sebastian L. and Mateo, Mario and Tolstoy, Eline},
   year={2008},
   month=apr, pages={L21–L24} }

@article{Gallart_2015,
   title={THE ACS LCID PROJECT: ON THE ORIGIN OF DWARF GALAXY TYPES—A MANIFESTATION OF THE HALO ASSEMBLY BIAS?},
   volume={811},
   ISSN={2041-8213},
   url={http://dx.doi.org/10.1088/2041-8205/811/2/L18},
   DOI={10.1088/2041-8205/811/2/l18},
   number={2},
   journal={The Astrophysical Journal},
   publisher={American Astronomical Society},
   author={Gallart, Carme and Monelli, Matteo and Mayer, Lucio and Aparicio, Antonio and Battaglia, Giuseppina and Bernard, Edouard J. and Cassisi, Santi and Cole, Andrew A. and Dolphin, Andrew E. and Drozdovsky, Igor and Hidalgo, Sebastian L. and Navarro, Julio F. and Salvadori, Stefania and Skillman, Evan D. and Stetson, Peter B. and Weisz, Daniel R.},
   year={2015},
   month=sep, pages={L18} }

@article{Sohn_2012,
   title={THE M31 VELOCITY VECTOR. I.HUBBLE SPACE TELESCOPEPROPER-MOTION MEASUREMENTS},
   volume={753},
   ISSN={1538-4357},
   url={http://dx.doi.org/10.1088/0004-637X/753/1/7},
   DOI={10.1088/0004-637x/753/1/7},
   number={1},
   journal={The Astrophysical Journal},
   publisher={American Astronomical Society},
   author={Sohn, Sangmo Tony and Anderson, Jay and van der Marel, Roeland P.},
   year={2012},
   month=jun, pages={7} }

@book{binney_tremaine,
  title     = "Galactic Dynamics: Second Edition",
  author    = "Binney, James  and Tremaine, Scott",
  year      = 2008,
  publisher = "Princeton University Press",
  address   = "Princeton, NJ"
}
